\documentclass[aip,apl,12pt,endfloats*]{revtex4-1}

\usepackage{graphicx}
\usepackage{xfrac}
\usepackage{amsmath} 
\newcommand{\angstrom}{\textup{\AA}}
\usepackage{textcomp}
\newcommand{\smallsim}{\smallsym{\mathrel}{\sim}}

\makeatletter
\newcommand{\smallsym}[2]{#1{\mathpalette\make@small@sym{#2}}}
\newcommand{\make@small@sym}[2]{%
  \vcenter{\hbox{$\m@th\downgrade@style#1#2$}}%
}
\newcommand{\downgrade@style}[1]{%
  \ifx#1\displaystyle\scriptstyle\else
    \ifx#1\textstyle\scriptstyle\else
      \scriptscriptstyle
  \fi\fi
}

\begin{document}

\title{Low-damping sub-10-nm thin films of lutetium iron garnet grown by molecular-beam epitaxy}

\author{C. L. Jermain}
\email{clj72@cornell.edu}
\affiliation{Cornell University, Ithaca, New York 14853, USA}
\author{H. Paik}
\affiliation{Cornell University, Ithaca, New York 14853, USA}
\author{S. V. Aradhya}
\affiliation{Cornell University, Ithaca, New York 14853, USA}
\author{R. A. Buhrman}
\affiliation{Cornell University, Ithaca, New York 14853, USA}
\author{D. G. Schlom}
\affiliation{Cornell University, Ithaca, New York 14853, USA}
\affiliation{Kavli Institute at Cornell for Nanoscale Science, Ithaca, New York 14853, USA}
\author{D. C. Ralph}
\affiliation{Cornell University, Ithaca, New York 14853, USA}
\affiliation{Kavli Institute at Cornell for Nanoscale Science, Ithaca, New York 14853, USA}

\date{\today}

\begin{abstract}
We analyze the structural and magnetic characteristics of (111)-oriented lutetium iron garnet (Lu$_3$Fe$_5$O$_{12}$) films grown by molecular-beam epitaxy, for films as thin as 2.8 nm. Thickness-dependent measurements of the in- and out-of-plane ferromagnetic resonance allow us to quantify the effects of two-magnon scattering, along with the surface anisotropy and the saturation magnetization. We achieve effective damping coefficients of $11.1(9) \times 10^{-4}$ for 5.3 nm films and $32(3) \times 10^{-4}$ for 2.8 nm films, among the lowest values reported to date for any insulating ferrimagnetic sample of comparable thickness. 
\end{abstract}


\maketitle 

Insulating ferrimagnets are of interest for spintronic applications because they can possess very small damping parameters, as low as $10^{-5}$ in the bulk.\cite{Sparks1964} They also provide the potential for improving the efficiency of magnetic manipulation using spin-orbit torques from heavy metals\cite{Miron2011,Liu2012} and topological insulators,\cite{Mellnik2014,Fan2014} because ferrimagnetic insulators will not shunt an applied charge current away from the material generating the spin-orbit torque. Making practical devices from ferrimagnetic insulators will require techniques capable of growing very thin films (a few tens of nm and below) while maintaining low damping. Much of the previous research in this field has focused on yttrium iron garnet (Y$_3$Fe$_5$O$_{12}$, YIG) grown by pulsed-laser deposition or off-axis sputtering,\cite{Kelly2013,Hahn2014,Montazeri2015,Hamadeh2014,Wang2014} but YIG is just one in a family of rare earth iron garnets with potentially useful properties.\cite{Geller1978} Here we examine the magnetic and structural properties of thin, (111)-oriented films of lutetium iron garnet (Lu$_3$Fe$_5$O$_{12}$, LuIG) grown by an alternative method, molecular-beam epitaxy (MBE).\cite{Schlom2008} We find that MBE is capable of providing sub-10-nm films with very low values of damping, rivaling or surpassing other deposition techniques. We are able to grow LuIG films down to 2.8~nm, or 4 layers along the interplanar spacing $d_{111}$ (0.71~nm),\cite{Kelly2012,Geller1978} while retaining high crystalline quality. We report in- and out-of-plane ferromagnetic resonance measurements as a function of film thickness, demonstrating reduced two-magnon scattering compared to previous work. We achieve effective damping coefficients as low as $11.1(9) \times 10^{-4}$ for 5.3 nm LuIG films and $32(3) \times 10^{-4}$ for 2.8 nm films, which can be compared to the best previous report for very thin YIG, $38 \times 10^{-4}$ for a 4 nm film.\cite{Kelly2013}

As an iron garnet, LuIG has ferrimagnetic properties similar to YIG. The magnetic moments in both materials arise from their Fe$^{3+}$ ions, which interact via super-exchange through oxygen atoms.\cite{Geller1978,Anderson1964} In bulk samples, LuIG has a slightly higher room-temperature saturation magnetization (1815 Oe) than YIG (1760 Oe).\cite{Geller1978,Anderson1964,VanUitert1971} The bulk lattice parameters for LuIG (12.283 \angstrom) and YIG (12.376 \angstrom) differ by 0.75\%.\cite{Geller1957,Espinosa1962} Both materials can be grown on isostructural gadolinium gallium garnet (Gd$_3$Ga$_5$O$_{12}$, GGG) substrates, which have a cubic lattice parameter of 12.383 \angstrom. The resulting mismatch causes biaxial tensile strain with a maximum value of 0.81\% and 0.07\% for LuIG and YIG, respectively. High-quality YIG films have been grown previously using off-axis sputter deposition\cite{Wang2013,Wang2014,Liu2014,Chang2014,Brangham2016} and pulsed-laser deposition (PLD).\cite{Kelly2013,Dorsey1993,Manuilov2009,Manuilov2010,Heinrich2011,Onbasli2014,Howe2015,Tang2016} The best reported damping values for thin YIG films grown by PLD to date include $2.3 \times 10^{-4}$ for a 20 nm film,\cite{Kelly2013} $3.2 \times 10^{-4}$ for a 10 nm film treated with a post-growth etching procedure,\cite{Sun2012} and $0.7 \times 10^{-4}$ for a 20 nm film treated with a post-growth high-temperature anneal.\cite{Hauser2016} For off-axis sputtering, the best reported values include $6.1 \times 10^{-4}$ for a 16 nm film,\cite{Brangham2016} $12.4 \times 10^{-4}$ for a 10.2 nm film,\cite{Liu2014} and $0.9 \times 10^{-4}$ for a 22 nm film with a post-growth high-temperature anneal.\cite{Chang2014} Previous measurements of films thinner than 10 nm recorded significant two-magnon scattering,\cite{Kelly2013,Liu2014} and much larger damping parameters of $38 \times 10^{-4}$ for a 4 nm film and $16 \times 10^{-4}$ for a 7 nm film.\cite{Kelly2013}

\begin{figure}
\includegraphics{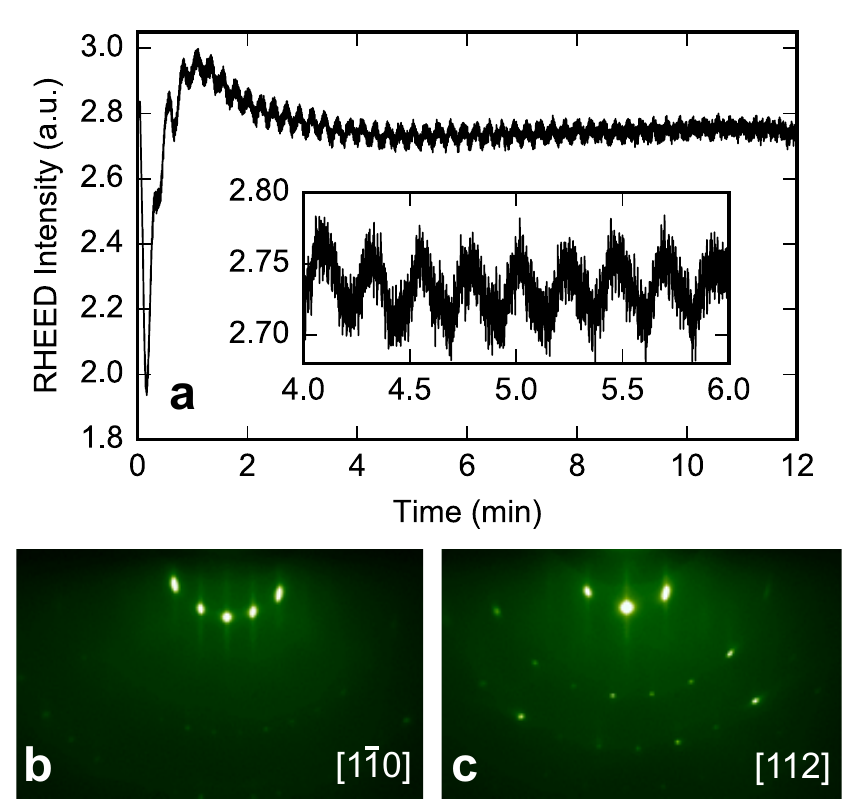}%
\caption{\label{figure-rheed}(a) RHEED intensity oscillations of a 10 nm thick LuIG film grown on a (111) GGG substrate, indicating layer-by-layer growth. Each oscillation peak-to-peak corresponds a single $d_{444}$ ($d_{111}/4$) spacing. (b,c) Kikuchi lines in the RHEED image taken along both $[1\bar{1}0]$ and $[11\bar{2}]$ azimuthal directions.}%
\end{figure}

Here we report the growth of epitaxial LuIG films with thicknesses from 2.8 to 40 nm by reactive MBE on (111) GGG substrates. (We study LuIG, rather than YIG, primarily because Lu is available within our MBE chamber.) Our substrates are prepared by annealing at 1300\textdegree C for 3 hr in an air furnace to produce well-defined unit-cell steps and smooth terraces (see Supplementary Information (SI)). During growth, we simultaneously co-supply Lu and Fe with an accuracy of $\pm$5\%, to achieve the stoichiometric atomic ratio of Lu:Fe=3:5. We use distilled ozone (O$_3$) at a background pressure of $1.0 \times 10^{-6}$ Torr as the oxidant. The growth temperature is 950 to 970\textdegree C, achieved by radiatively heating the backside of the GGG substrates, which are coated with 400 nm of Pt to enhance thermal absorption. The quality of crystal growth is monitored using \textit{in-situ} reflection high-energy electron diffraction (RHEED) along both the $[1\bar{1}0]$ and $[11\bar{2}]$ in-plane azimuthal directions. The RHEED intensity oscillations (Fig.~\ref{figure-rheed}(a)) indicate layer-by-layer growth,\cite{Ichimiya2004} with an oscillation period corresponding to the $d_{444}$ spacing, which is a quarter of a single LuIG layer ($d_{111} = 0.71$~nm) along the (111)-orientation. We also observe sharp RHEED features and clear Kikuchi lines during growth, as seen in Fig.~\ref{figure-rheed}(b,c) for a 10 nm film, demonstrating that our films are of high crystalline quality. These features are not observed if the flux drifts more than $\pm$5\%, or if the growth temperature is less than 900\textdegree C.

\begin{figure}
\includegraphics{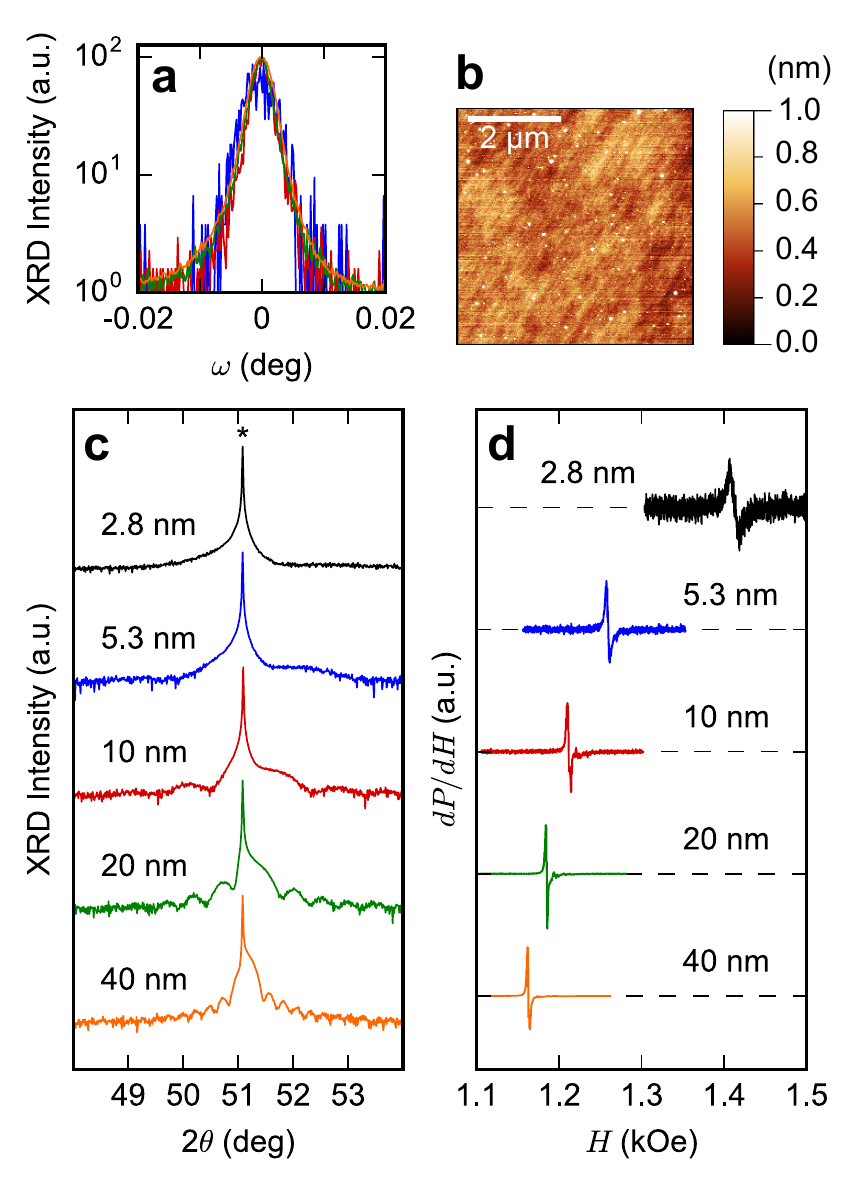}%
\caption{\label{figure-panel}(a) X-ray diffraction (XRD) rocking curves for all of the LuIG thicknesses above 2.8 nm indicate commensurate growth and consistent strain. (b) Representative atomic force microscopy image of the surface of the 2.8 nm film, showing a RMS roughness of 0.26 nm over 5 \textmu m x 5 \textmu m scan size, which indicates the roughness is substrate limited. (c) $\theta/2\theta$ XRD scans of LuIG thin films grown on (111) GGG substrates as a function of film thickness. The asterisk marks the 444 GGG substrate reflection. (d) Normalized derivative-absorption FMR spectra of the corresponding samples taken at 5 GHz show narrow linewidths that decrease for increasing thickness. The resonance position also depends on the thickness.}%
\end{figure}

We quantify the strain state and verify the crystalline quality with four-circle X-ray diffraction (XRD) measurements. The normalized rocking curves for films with different thicknesses (except the 2.8 nm film), overlaid in Fig.~\ref{figure-panel}(a), all have full-width at half-maximum (FWHM) values that are less than 0.004\textdegree, limited by the GGG substrate. This indicates that our films are commensurately strained, and are at the maximal strain state of 0.81\% set by the lattice mismatch with the substrate. While the rocking curve measurements on the 2.8 nm film lack sufficient signal-to-noise for analysis, the thicker films suggest that the strain state is also commensurate for this film. The surfaces of the films are characterized by atomic force microscopy. Figure~\ref{figure-panel}(b) shows the 2.8 nm film, with a measured surface roughness of 0.26 nm (RMS) over a 5 \textmu m x 5 \textmu m scan area. This indicates that the surface quality is substrate limited, which we observe for all thicknesses. Figure~\ref{figure-panel}(c) shows the $\theta/2\theta$ XRD patterns of the LuIG thin films for all thicknesses grown. The visible Laue oscillations confirm thickness measurements we make with the RHEED intensity oscillations and flux calibrations. Low-angle X-ray reflectively (XRR) determines the film thicknesses as 2.84(1), 5.33(2), 9.94(2), 20.16(3) and 40.37(10) nm, which we nominally report as 2.8, 5.3, 10, 20, and 40 nm.

The magnetic properties of the MBE-grown LuIG films are characterized by measuring the frequency and thickness dependence of ferromagnetic resonance (FMR). The samples are placed, LuIG-side down, on a broadband coplanar waveguide so that the Oersted field of the waveguide excites FMR at GHz frequencies.\cite{Kalarickal2006} We measure the FMR spectra at fixed frequency by sweeping the applied magnetic field, oriented either in-plane (IP) parallel to the coplanar waveguide or out-of-plane (OOP). For the IP measurements, we position the film so that the applied magnetic field is always along the $[11\bar{2}]$ crystal orientation. The measured signal corresponds to the derivative absorption, which we detect via the voltage from a detector diode. We achieve optimal sensitivity using lock-in amplification by modulating both the input power and the applied field. All of the FMR measurements are performed at room temperature. Further details of the FMR apparatus are described in the SI.

\begin{figure}
\includegraphics{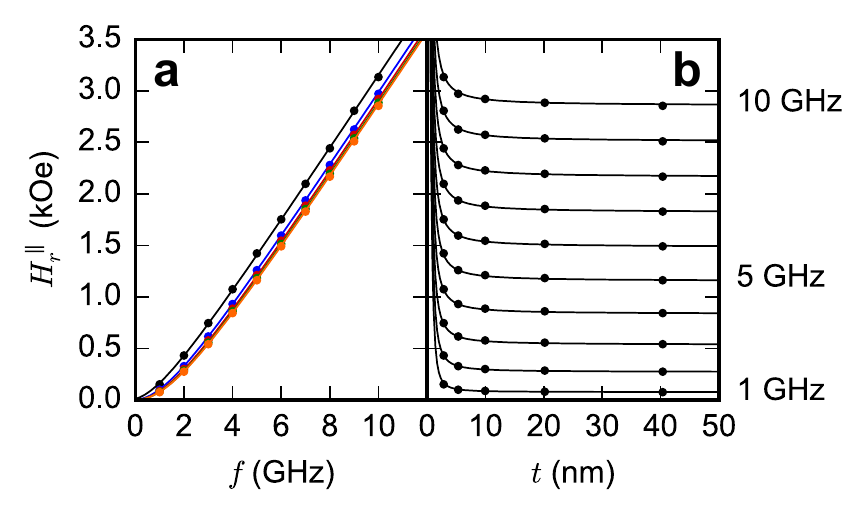}%
\caption{\label{figure-resonance}(a,b) In-plane FMR resonance fields of each LuIG sample (a) as a function of frequency for different sample thicknesses and (b) as a function of thickness for different frequencies. The solid lines in (a) and (b) represent simultaneous fits to Eq.~(\ref{equation-resonance}) with the 3 fitting parameters $r$, $4\pi M_s$, and $K_s$.}%
\end{figure}

Figure~\ref{figure-panel}(d) shows the IP-FMR response at 5 GHz for LuIG samples with different thicknesses. Two trends are apparent as the film thickness is reduced: (i) the resonance position shifts to higher fields and (ii) the linewidth increases substantially. Below we show that both of these effects can be explained by two-magnon scattering.\cite{Arias1999,Mills2003,Lenz2006} We focus first on the behavior of the resonance fields. We have measured the IP-FMR resonances for each film thickness at frequencies from 1 to 10 GHz. The evolution as a function of frequency is shown in Fig.~\ref{figure-resonance}(a) and as a function of thickness in Fig.~\ref{figure-resonance}(b).

In the presence of two-magnon scattering, the IP resonance field $H_r^\parallel$ predicted by the Kittel equation in the thin-film limit takes the form\cite{Arias1999,Azevedo2000}

\begin{equation}
\label{equation-resonance}
\begin{split}
H_r^\parallel(f,t)& = \sqrt{\left(\frac{4\pi M_\mathrm{eff}(t)}{2}\right)^2 - \left(\frac{2\pi f}{|\gamma|} + \Delta H_r(t)\right)^2} \\
&\quad - \frac{4\pi M_\mathrm{eff}(t)}{2},
\end{split}
\end{equation}

\noindent with $f$ the excitation frequency, $t$ the film thickness, $\Delta H_r$ a renormalization shift associated with two-magnon scattering, and $\gamma$ the gyromagnetic ratio. We measured $|\gamma|/2\pi = 2.77(2)$ MHz/Oe based on the frequency dependence of the OOP resonance field $H_r^\perp$ (see SI). The effective anisotropy field $4\pi M_\mathrm{eff}$ is expected to depend on the film thickness, because it contains contributions from both bulk demagnetization and surface anisotropy: 

\begin{equation}
\label{equation-meff}
4\pi M_\mathrm{eff} = 4\pi M_s + \frac{2 K_s}{M_s t}.
\end{equation}

\noindent Here $M_s$ is the saturation magnetization and $K_s$ is the surface anisotropy energy. The renormalization shift produced by two-magnon scattering can be related to the surface anisotropy as\cite{Arias1999,Azevedo2000}

\begin{equation}
\label{equation-shift}
\Delta H_r(t) = r \left( \frac{2 K_s}{M_s t} \right)^2,
\end{equation}

\noindent where $r$ is a parameter characterizing the strength of two-magnon scattering. 

We performed a global least-squares fit of Eqs.~(\ref{equation-resonance})-(\ref{equation-shift}) to all the data in Fig.~\ref{figure-resonance} using three fitting parameters $r$, $4\pi M_s$, and $K_s$. As shown by the lines in Fig.~\ref{figure-resonance}, we find excellent fits assuming that all three parameters are independent of film thickness, obtaining the values $r = 4.9(2) \times 10^{-4}$ Oe$^{-1}$, $4\pi M_s = 1609(1)$ Oe, and $K_s = -8.52(8) \times 10^{-3}$ erg/cm$^2$. We also attempted to fit the data without the two-magnon contribution (i.e., with the constraint $r=0$ Oe$^{-1}$), but we found significant discrepancies for the 2.8 film, especially at low frequencies (see SI). The non-zero value of $r$ implies that the two-magnon mechanism is active. For our 2.8 nm film, the renormalization shift is $\Delta H_r = 110$ Oe, similar to that found in a 2.7 nm NiFe film.\cite{Azevedo2000} This is the first report of the renormalization shift in iron garnets. The value of $4\pi M_s$ determined by the fit is significantly lower than the bulk LuIG value of 1815 Oe.\cite{VanUitert1971} This reduction is qualitatively consistent with the tensile strain in our films from the GGG substrate. The tensile strain is expected to enhance the antiferromagnetic super-exchange interaction between the two inequivalent Fe$^{3+}$ lattices in the LuIG and therefore reduces the overall saturation magnetization.\cite{Geller1978,Anderson1964} The negative sign that we find for $K_s$ indicates that the surface anisotropy reduces the effective demagnetization field $4\pi M_\mathrm{eff}$ compared to the bulk value. The magnitude of $K_s$ is relatively weak, however (e.g., more than two orders of magnitude smaller than $K_s$ for annealed CoFeB).\cite{Worledge2011} With our values for $4\pi M_s$ and $K_s$, only for extremely thin LuIG films, $< 0.8$ nm, might the magnetic anisotropy be turned perpendicular to the sample plane. For any thickness above this, $4\pi M_\mathrm{eff}$ favors in-plane magnetization.

Next we consider the FWHM linewidths ($\Delta H$) of the IP FMR resonances for our LuIG films as a function of thickness and FMR frequency. The linewidths of our samples are sufficiently narrow that small inhomogeneities in the films can result in overlapping but distinguishable resonances, as has often been seen previously in measurements on thin garnet films.\cite{Jalali2002,Heinrich2011,Kelly2013} To make an accurate determination of the intrinsic linewidths, we fit each measured curve to the sum of multiple (2 in this analysis) Lorentzian derivative curves with their widths constrained to be identical (see SI for details). This procedure produces values for the linewidth that are consistent with the results for films that can be cleaved into samples sufficiently small to isolate a single resonance (see SI).

\begin{figure}
\includegraphics{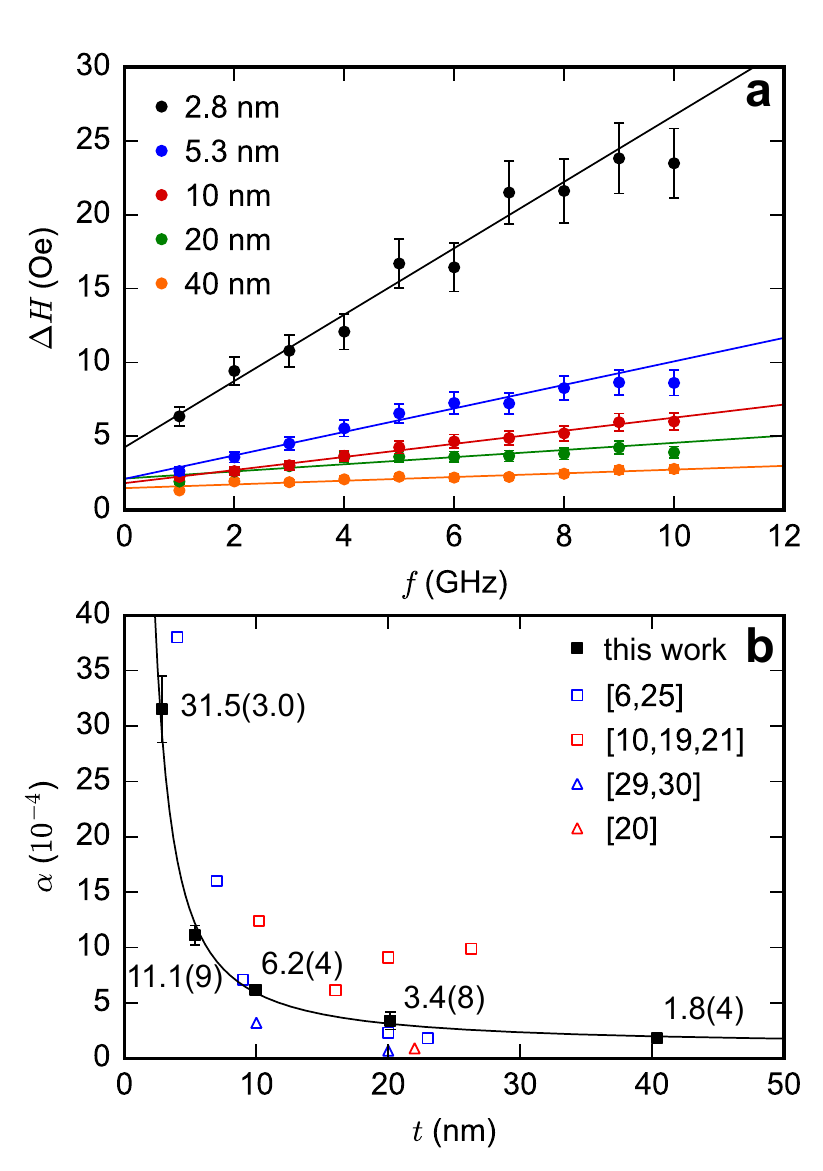}%
\caption{\label{figure-linewidth}(a) Frequency dependence of the FMR linewidth, for LuIG films of different thickness. The linewidths are fit to straight lines up to 8 GHz, after which the linewidths start to roll off, following the signature of two-magnon scattering. (b) Thickness dependence our measured values of magnetic damping (black squares). The line depicts the phenomenological form of Eq.~(\ref{equation-damping}). Previously-reported results for damping in thin YIG films are shown for films deposited by PLD (open blue symbols) PLD and off-axis sputtering (open red symbols). Open triangles represent post-processed films.}%
\end{figure}

Figure~\ref{figure-linewidth}(a) shows the measured frequency dependence of the linewidth for each of our films. We observe a linear dependence on frequency up to $\smallsim$8 GHz. At higher frequencies, the linewidths deviate from linearity, most obviously for the 2.8 and 5.3 nm films. This high-frequency curvature is qualitatively consistent with the effect of two-magnon scattering, as observed previously in PLD-grown YIG films.\cite{Kelly2013} Using the expression\cite{Kalarickal2006}

\begin{equation}
\label{equation-linewidth}
\Delta H(f) = \frac{4 \pi \alpha f}{|\gamma|} + \Delta H_0,
\end{equation}

\noindent we can define an effective Gilbert damping parameter, $\alpha$, for each value of film thickness based on linear fits to the data below 8 GHz (Fig.~\ref{figure-linewidth}(b)). The line shown in Fig.~\ref{figure-linewidth}(b) is a fit to a phenomenological form

\begin{equation}
\label{equation-damping}
\alpha = \alpha_G + \alpha_{2M} \left( \frac{A}{t^2} + \frac{B}{t} \right),
\end{equation}

\noindent with $\alpha_G = 0.9(6) \times 10^{-4}$, $A\alpha_{2M} = 125(45) \times 10^{-4}$ nm$^2$, and $B\alpha_{2M} = 36(11) \times 10^{-4}$ nm. 

Our damping values are among the best reported for any garnet film, and for the first time extend the viable thickness of low-damping thin films well below 10 nm. We measure $\alpha = 11.1(9) \times 10^{-4}$ for 5.3 nm LuIG films and $32(3) \times 10^{-4}$ for 2.8 nm films. We speculate that our MBE growth procedure minimizes the amount of surface roughness and other defects even for very thin LuIG films, compared to other deposition techniques, and thereby provides a reduced level of two-magnon scattering. Similar MBE growth procedures may also allow the production of sub-10-nm films made from YIG and other garnets, assisting in the development of a wide variety of spintronic devices incorporating these materials.

\begin{acknowledgments}
We acknowledge F.~Guo for helpful discussion on two-magnon theory. This work was supported by the National Science Foundation (DMR-1406333) with partial support from the Cornell Center for Materials Research (CCMR), part of the NSF MRSEC program (DMR-1120296). We made use of the Cornell Nanoscale Facility, a member of the National Nanotechnology Coordinated Infrastructure (NNCI), which is supported by the NSF (ECCS-1542081) and also the CCMR Shared Facilities. The work of H.P.~and D.G.S.~is supported by the Air Force Office of Scientific Research under award number FA9550-16-1-0192.
\end{acknowledgments}

\bibliography{luig.bib}

\end{document}